# Measurement of the transverse diffusion coefficient of charge in liquid xenon

W.-T. Chen[a], H. Carduner[b], J.-P. Cussonneau[c], J. Donnard[d], S. Duval[e],
A.-F. Mohamad-Hadi[f], J. Lamblin[g], O. Lemaire[h], P. Le Ray[i], E. Morteau[j],
T. Oger[k], L. Scotto-Lavina[l], J.-S. Stutzmann[m], and D. Thers[n]

Subatech, Ecole des Mines, CNRS/IN2P3 and Université de Nantes, 44307 Nantes, France

[a]nchen@subatech.in2p3.fr, [b]Herve.Carduner@subatech.in2p3.fr,
[c]Jean-Pierre.Cussonneau@subatech.in2p3.fr, [d]Jerome.Donnard@subatech.in2p3.fr,
[e]Samuel.Duval@subatech.in2p3.fr, [f]Abdul-Fattah.Mohamad-Hadi@subatech.in2p3.fr,
[g]Jacob.Lamblin@subatech.in2p3.fr, [h]olivier.lemaire@subatech.in2p3.fr,
[i]Patrick.Leray@subatech.in2p3.fr, [j]Eric.Morteau@subatech.in2p3.fr,
[k]tugdual.oger@subatech.in2p3.fr, [l]scotto@subatech.in2p3.fr,
[m]Jean-Sebastien.Stutzmann@subatech.in2p3.fr, [n]Dominique.Thers@subatech.in2p3.fr

**Keywords:** diffusion coefficient, noble-liquid detector (scintillation and ionization), photon detectors, Compton telescope, xenon, medical imaging

**Abstract**

Liquid xenon (LXe) is a very attractive material as a detection medium for ionization detectors due to its high density, high atomic number, and low energy required to produce electron-ion pairs. Therefore it has been used in several applications, like γ detection or direct detection of dark matter. Now Subatech is working on the R & D of LXe Compton telescope for 3γ medical imaging, which can make precise tridimensional localization of a ($\beta^+$, γ) radioisotope emitter. The diffusion of charge carriers will directly affect the spatial resolution of LXe ionization signal. We will report how we measure the transverse diffusion coefficient for different electric field (0.5 ~ 1.2 kV/cm) by observing the spray of charge carriers on drift length varying until 12cm. With very-low-noise front-end electronics and complete Monte-Carlo simulation of the experiment, the values of transverse diffusion coefficient are measured precisely.

**Introduction**

Liquid xenon (LXe) is an attractive material as radiation detector medium due to its high atomic number (54) and high density ( 3 g/cm$^3$), which makes it very efficient to stop penetrating radiation [1]. To be a scintillator material, LXe not only provides high stopping power benefit in a single large and homogeneous volume, but also gives a fast decay time of the scintillation signal. For the detection of low energy particle (< 1 MeV) by measuring the ionization yield, LXe has a small W-value (15.6 eV, which is the average energy required to produce an ion pair), hence it can provide a large ionization yield.

Twenty years ago the standard size of LXe detectors was still limited to less than 1 kg of mass. Thanks to the development of technology in these years for cryocooler, purifier, charge read-out sensors, and the photodetectors with high quantum efficiency at 178 nm (wavelength of LXe scintillation), today there are several experiments which use more than 100 kg of LXe. The most famous experiment is XENON Dark matter search, which is a phased program aiming at progressively improved sensitivity by the series of two-phase time projection chambers (TPCs): XENON10, XENON100, and XENON1T, the numbers referring to the order of magnitude of fiducial target mass in kg/ton. Figure 1 presents the principle of a typical two-phase TPC, developed for the detection of dark matter weakly interacting massive particles (WIMPs). Both the scintillation

(S1) and ionization (S2), produced by radiation in the sensitive LXe volume are detected. Ionization electrons drift toward the anode due to an applied electric field but once they reach the liquid surface, they are extracted into the gas phase where they emit proportional scintillation light. Both the direct scintillation (S1) and the proportional scintillation (S2) are detected with Vacuum Ultraviolet (VUV) sensitive photomultiplier tubes (PMTs), located in the gas and in the liquid. The (x,y) event localization is provided by the PMTs in the gas by detecting the S2; the third coordinate, along the drift direction, is inferred from the time difference between S1 and S2 and the known drift velocity in the liquid. The ratio of S2 yield and S1 yield is used to distinguish the WIMP events and γ/electron events. In 2011 XENON Collaboration has presented results from the direct search for dark matter with the XENON100 detector, which leads to the most stringent limit on dark matter interactions today [2]. XENON1T, the next step of XENON project, which will use 2.4 ton of xenon, is already under construction and will start data taking before 2015.

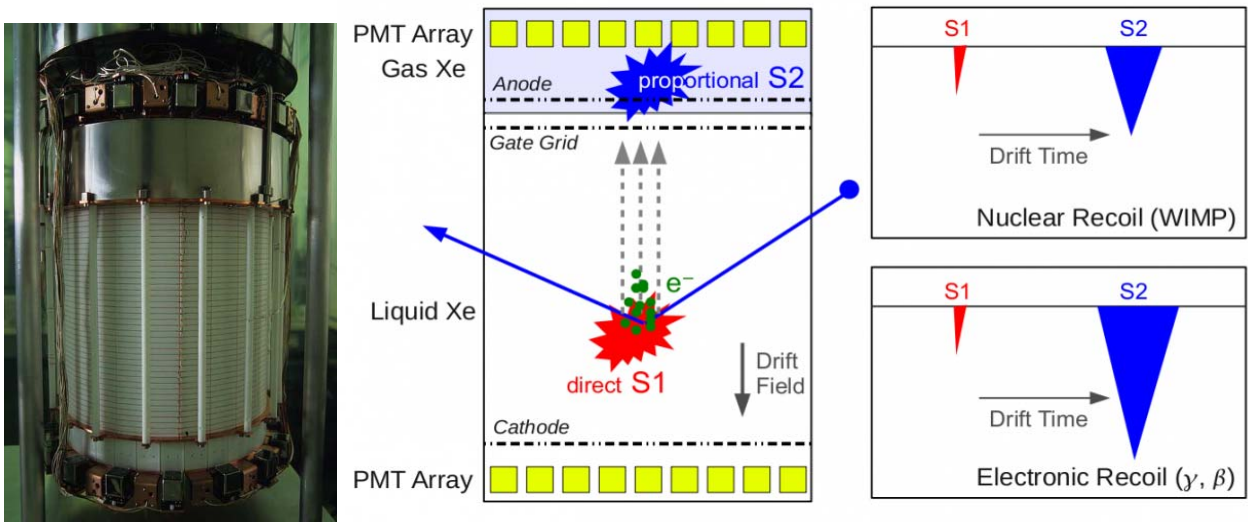

Figure 1: (left) the two-phase xenon TPC used in XENON100; (right) the principle of a typical two-phase TPC, developed for the detection of WIMPs. The nuclear recoil (WIMP) events and the electronic recoil (background from γ or β) can be distinguished by measuring the ratio of S2 and S1 amplitude.

The excellent properties of LXe for γ detection make it also attractive for medical imaging. Few years ago, a novel imaging technique was initiated at Subatech [3], which proposes a 3γ medical imaging using a specific radioisotope emitting a γ just after its $β^+$ decay. A single-phase TPC is used to detect the additional γ-ray: the direction is inferred from the Compton kinematics as shown in Figure 2. The first hit energy leads to the determination of the aperture angle θ of the Compton cone and the position of the emitter can then be measured by calculating the intersection between the cone and the line-of-response (LOR), which is measured by a classical PET device. This technology permits an event-by-event detection, which means that high resolution of image can be reached with much less radiation dose.

To make sure the idea of 3γ imaging can work well, the detector must be able to provide good spatial and energy resolution. The spatial resolution of interaction position is physically limited by the spray of charge cloud ($\sigma/\sqrt{Z} = \sqrt{2D_T v_d}$, where $D_T$, $Z$ and $v_d$ are the transverse diffusion coefficient, drift length, and drift velocity, respectively.), which is also related to the detector design to get the best energy resolution. However, there are very few publications which present the measurement of transverse diffusion coefficient [4-5]. In this paper, the measurement of the charge transfer in LXe is presented; the drift velocity and transverse diffusion coefficient of electrons are well measured with high precision.

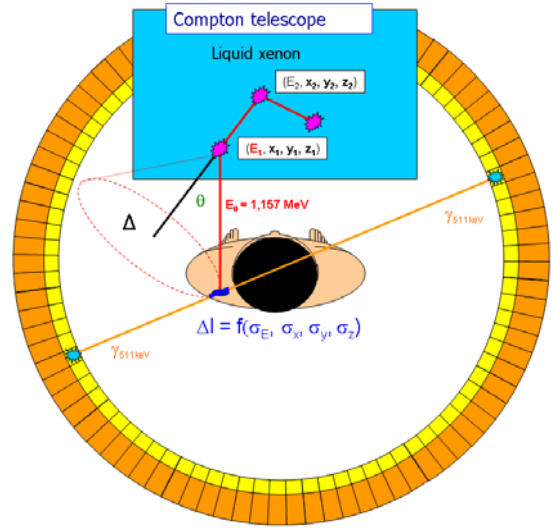

Figure 2: The principle of 3 γ medical imaging. The position of the radioisotope can be located by calculating the intersection between the cone and the LOR.

**Apparatus and procedure**

The XEnon Medical Imaging System (XEMIS), which is a liquid-xenon time-projection chamber, has been built in order to validate the 3γ imaging concept. The schema of the whole cryogenic system is presented in Figure 3. A pulse tube refrigerator (PTR), which is optimized for the use of xenon liquefaction and can provide 100 W cooling power at 165 K , is not only used to liquefy the xenon to liquid phase but also to keep it around 171 K under a pressure of 1.2 bar. The distance between the cryostat and the PTR is extended to ~2m to reduce mechanical noise. A purification circuit equipped with a plate heat exchanger, an oil-free membrane pump, a flow controller, and a getter, is used to purify the xenon, since the xenon is required to be as pure as possible to prevent charge carriers capture by electronegative impurities ($H_2O$, $O_2$) released by construction materials into the liquid [6]. The heat loss of the whole system is ~46 W, and the flow rate of the purification can still reach ~4.2 L/min because the heat exchanger can recuperate some of the vapor heat of liquid to liquefy the gas. Figure 4 presents the structure of the TPC (length: 12 cm, diameter: 3.6 cm) which is composed of a PMT for the detection of VUV scintillation photons, field shaping rings to provide drift electric field, a MICROMEGAS device for charge carriers readout, and a 2.54 cm × 2.54 cm anode segmented to 16 pixels. An ultra-low noise front-end electronics – IdeF-X [7], which has 16 channels, is used for the charge readout. The noise level on each channel was measured to be ~100 electrons in 171 K, which contributes less than 0.5% of energy resolution for 511 keV energy deposit. A collimated $^{22}$Na ($E_{max}$ $β^+$= 545 keV, $E_γ$ = 1.257 MeV) source was placed in front of the TPC. Two 511 keV back-to-back γ-rays can trigger the chamber in coincidence with a CsI crystal coupled to a PMT at the opposed side.

Measurement of electron attenuation length is done by recording the energy spectrum corresponding to the 511 keV γ-rays interacting at different depths of the sensitive volume. All charge transfer measurements are performed when attenuation length is greater than 15 cm.

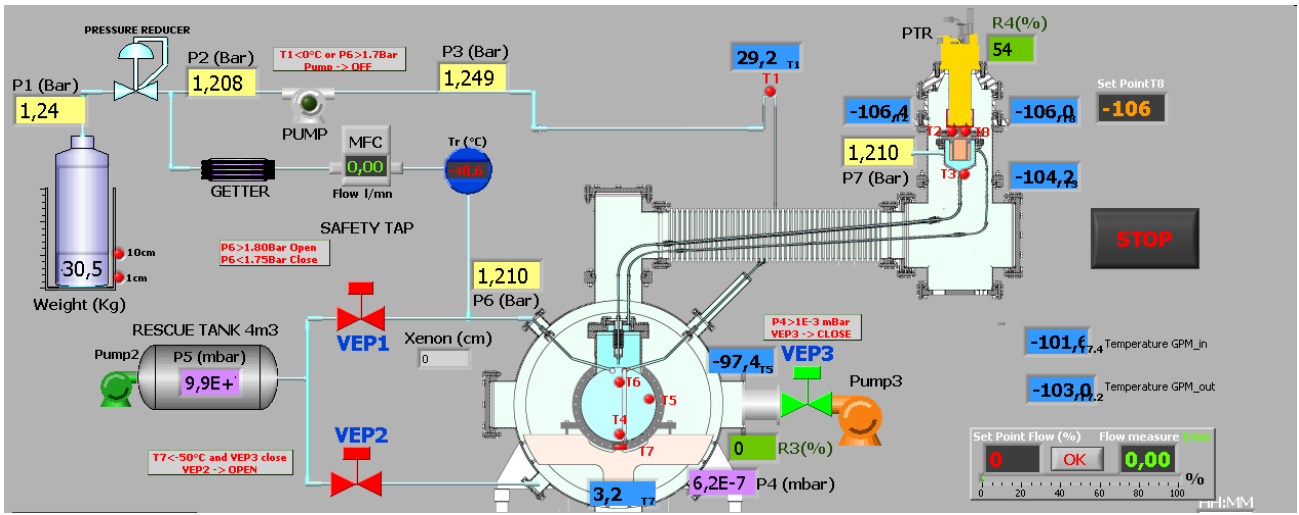

Figure 3: The diagram of the cryogenic system of XEMIS.

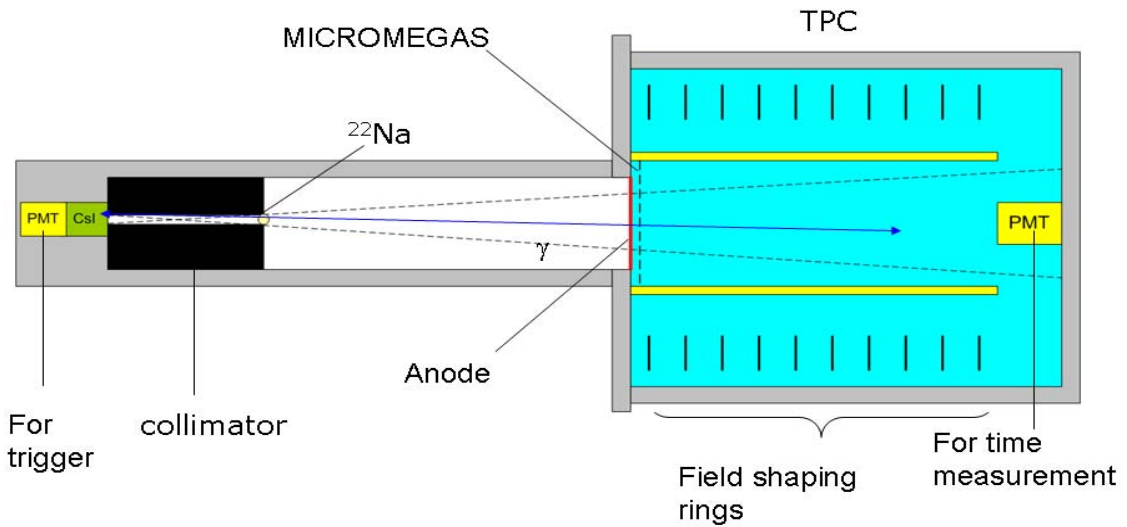

Figure 4: Schematic drawing of the coincidence setup used for 511 keV γ-rays selection.

**Results**

The collected charge of 511 keV γ-ray energy deposit as a function of drift electric field (**E**) is presented in Figure 5 (left), which is obtained by the collected charge after the correction of attenuation. The related yield is normalized to the charge yield at **E**=1.0 kV/cm. The result indicates that the increase of the field strength can reduce the probability of the recombination of ion pair so more electrons can drift to the anode, which is consistent with other measurements [8].

The charge drift velocity ($v_d$) as a function of **E** is presented in Figure 5 (right), which is estimated from the ratio of the maximum drift length (12 cm) and the maximum drift time. The time resolution of the charge readout is 80 ns, so the drift velocity can be measured precisely. The result is also consistent with other measurement [9]. With our current setup and the drift velocity we measured, 0.5 mm of z resolution can be reached.

For the measurement of transverse diffusion coefficient, the cluster multiplicity of events as a function of drift length, which will increase due to the increase of charge spray, is measured. A series of Monte-Carlo (MC) simulation using the information of XEMIS geometry, noise level, charge yield, and attenuation length are prepared with different charge spray ($\sigma/\sqrt{Z}$). Hence $\sigma/\sqrt{Z}$ of data can be estimated from the variation of cluster multiplicity as a function of Z, which is shown in Figure 6 (left). The $\sigma/\sqrt{Z}$ as a function of electric field is measured and presented in Figure 6 (middle). With the relation between charge spray of transverse diffusion coefficient ($D_T$): $D_T = \left(\sigma/\sqrt{Z}\right)^2 v_d/2$, where $v_d$ is also measured in this work, $D_T$ is estimated as a function of the ratio of drift electric field and atomic density in liquid xenon. The result is consistent with the existed measurement [1, 4-5].

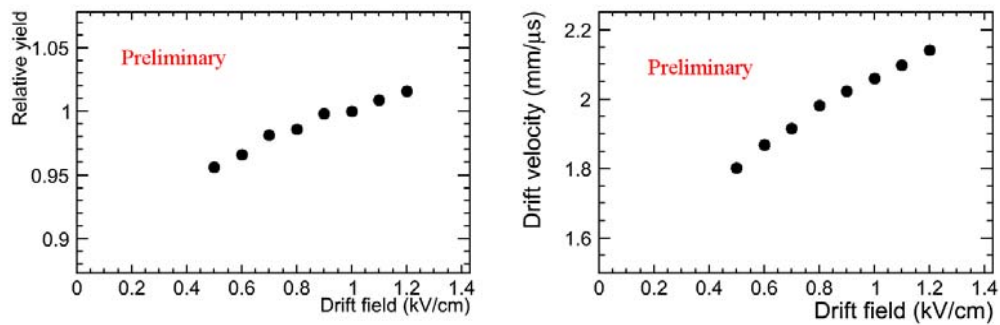

Figure 5: (left) charge yield on LXe normalized to the one at E=1.0 kV/cm as a function of drift field. (right) electron drift velocity as a function of the drift field in LXe. Both results are measured in 171 K LXe (1.2 bar).

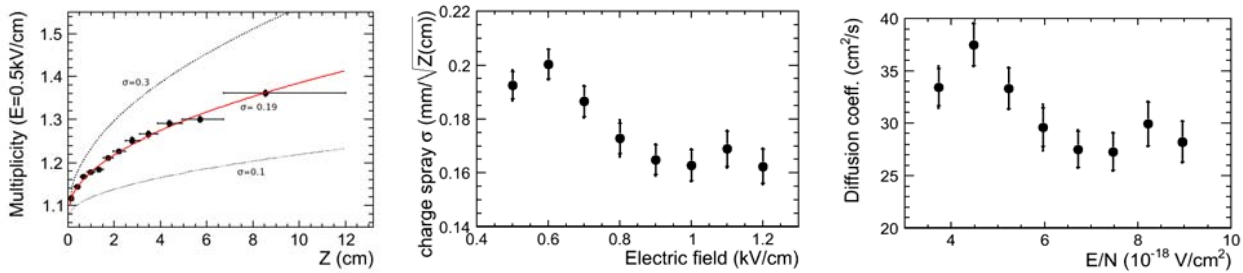

Figure 6: (left) the cluster multiplicity as a funtion of Z, which is measured at E=0.5 kV/cm. The shape is consistent with the MC simulation with charge spray ($\sigma/\sqrt{Z}$) = 0.19 mm. (middle) $\sigma/\sqrt{Z}$ as a function of drift electric field. (right) The transverse diffusion coefficient as a function of the ratio of electric field and atomic density. All results are obtained in 171K LXe (1.2 bar).

**Summary**

Thanks to the development of necessary technology in these years, today LXe detectors can be used in lots of applications with excellent performance. To use large-size LXe detectors in the future, the properties of charge transfer in the liquid must be well understood.

With the ultra-low-noise electronics and precise time measurement, related collected charge, drift velocity, charge spray $\sigma/\sqrt{Z}$, and transverse diffusion coefficient of electrons in LXe as a function of drift electric field (0.5-1.2 kV/cm) were measured in the temperature 171 K with high precision. The measured values of $\sigma/\sqrt{Z}$ are very promising and closed to 0.17 mm at 1 kV/cm. It is a strong input to consider in view of being able to reach spatial resolution better than 0.1 mm with a liquid xenon Compton telescope.

**Acknowledgements**

This work is supported by the region of Pays de la Loire, France.